\documentclass[twocolumn,showpacs,preprintnumbers,amsmath,amssymb]{revtex4}
\usepackage{graphicx}% Include figure files
\usepackage{dcolumn}% Align table columns on decimal point
\usepackage{bm}% bold math

\begin{document}

\title{Charge ordering and coexistence of charge fluctuations in\\
quasi-two-dimensional organic conductors $\theta$-(BEDT-TTF)$_2X$}% Force line breaks with \\

\author{Masafumi Udagawa}
\author{Yukitoshi Motome}%
\affiliation{%
Department of Applied Physics, 
University of Tokyo, 
Tokyo 113-8656, Japan}%

\date{\today}% It is always \today, today,

\begin{abstract}
We present a scenario for the peculiar coexistence of charge fluctuations
observed in quasi-2D $1/4$-filled organic conductors
$\theta$-(BEDT-TTF)$_2X$ in the quantum critical regime 
where the charge ordering is suppressed down to zero temperature. 
The scenario is explored in the extended Hubbard model 
including electron-phonon couplings 
on an anisotropic triangular lattice. 
We find that the coexisting fluctuations emerge from 
two different instabilities, the ``Wigner crystallization on lattice'' 
driven by the off-site Coulomb repulsion and
the charge-density-wave formation due to the nesting of the Fermi surface,
not from phase competition or real-space inhomogeneity. 
This mechanism explains the contrastive temperature dependence of 
two fluctuations in experiments. 
\end{abstract}

\maketitle

Charge ordering (CO) is a periodic arrangement of electrons 
which drives a metal-insulator transition. 
It is often found in organic conductors as well as transition metal compounds,
and it plays a key role in electronic properties such as transport phenomena.
Careful control of chemical substitutions and/or the external pressure
has revealed a systematic change of the electronic states, and has
opened up the possibility of a comprehensive understanding of 
the correlated electron systems
\cite{rf:reviews,rf:reviews2}.

\begin{figure}[b]
\begin{center}
\includegraphics[width=0.34\textwidth]{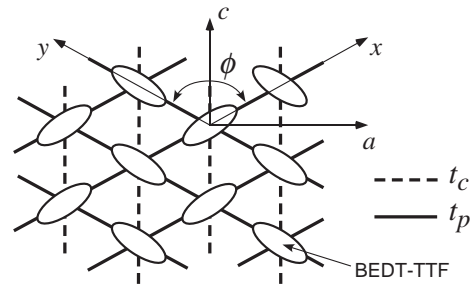}
\end{center}
\caption{\label{lattice} 
Schematic picture of BEDT-TTF layer.
Ellipses represent BEDT-TTF molecules, which form an anisotropic triangular lattice. 
$\phi$ is the dihedral angle, and
$t_p$ and $t_c$ are the transfer integrals for the nearest-neighbor molecules
used in the tight-binding Hamiltonian.
See the text for details.}
\end{figure}

Among them, quasi-2D organic conductors $\theta$-(ET)$_2X$ are
intriguing systems, where ET is an abbreviation for BEDT-TTF and X represents
monovalent closed-shell unit.
The crystal structure consists of 
an alternating stack of ET and $X$ layers; 
ET molecules constitute conducting layers of 
an anisotropic triangular structure 
as schematically shown in Fig.~\ref{lattice}, and 
anions $X^-$ form insulating layers. 
The charge transfer between ET and $X^-$ layers
leads to hole doping at $1/4$ filling in the ET layers 
(one hole per two ET molecules on average).
By changing $X$ units and/or by applying pressure,
the compounds exhibit a rich variety of electronic phases at low temperatures ($T$),
such as CO insulator, spin-Peierls state, and superconductivity.
It was proposed that 
the variety is summarized in a systematic way
by using the angle between neighboring ET molecules,
i.e., the dihedral angle $\phi$ (Fig.~\ref{lattice}) 
\cite{rf:MoriH98_1,rf:MoriH98_2}.
It is considered that $\phi$ modifies the transfer integrals 
between ET molecules, $t_p$ and $t_c$, 
in particular, the ratio between them, $t_c/t_p$.
In fact, in the series of $X$ = $MM'$(SCN)$_4$ with $M'$ = Co or Zn,
a stripe-type CO takes place in compounds with $M$ = Tl and Rb
\cite{rf:Miyagawa00,rf:Chiba01,rf:WatanabeM03,rf:WatanabeM04}, 
and the transition temperature is decreased 
from $T_{\text{CO}} = 250$~K for Tl to $190$~K for Rb,
as $\phi$ decreases from $\phi=116^\circ$ to $111^\circ$
and correspondingly $t_c/t_p$ decreases from $0.5$ to $0.4$.
Finally, $T_{\text{CO}}$ is suppressed down to the lowest $T$ for $M$ = Cs
with $\phi=104^\circ$ and $t_c/t_p \sim 0.1$.

The marginal compound with $M$ = Cs, close to the quantum critical point
where $T_{\text{CO}} \to 0$, 
has attracted special attention 
because of its peculiar low-$T$ properties.
A striking feature was found in X-ray scattering results:
two diffuse peaks were observed at different wave numbers
\cite{rf:WatanabeM99,rf:Nogami99}.
One appears at the same wave number as that of the stripe-type CO in Tl and Rb compounds,
which has a twofold period along the $c$ axis.
The other peak corresponds to a threefold period both in the $a$ and $c$ axes.
It is remarkable that charge fluctuations with different wave numbers coexist.
Moreover, these diffuse peaks exhibit contrastive $T$ dependence:
The peak for the twofold period grows rapidly below $T \sim 90$~K, 
while the threefold one shows little $T$ dependence
\cite{rf:WatanabeM99,rf:Nogami99}.
Recently, anomalous transport properties were also observed 
\cite{rf:Inagaki04,rf:Sawano05,rf:Yamaguchi06},
and their relation to the characteristic charge fluctuations has been issued.
In fact, it was found that
the peak for the twofold period is suppressed under the electric current, 
while the threefold one is not
\cite{rf:Sawano05}.

A scenario to understand the peculiar behaviors is phase competition
among different CO states. 
Several different CO states are stabilized due to strong Coulomb repulsion
\cite{rf:Wigner34,rf:Seo00,rf:MoriT00,rf:MoriT03}, 
which are sometimes considered as generalization of ``Wigner crystals''
\cite{rf:Hubbard78}.
It was shown that, in a certain parameter regime, 
some of the CO states are energetically degenerate, 
and a keen competition among them may suppress long-range CO and
even lead to metallic behavior 
\cite{rf:Merino05,rf:Kaneko06,rf:WatanabeH06,rf:Hotta06_1,rf:Hotta06_3}.
It was speculated that the phase competition leads to 
an inhomogeneous state 
where the competing phases are mixed in the real space 
by forming domainlike structure
\cite{rf:Inagaki04,rf:Sawano05}.
However, it is not obvious how such inhomogeneity emerges 
in the clean system almost free from impurities 
and why different domains grow with the contrastive $T$ dependence.
Furthermore, there is no experimental evidence for the domainlike state so far. 

In this letter, we present an alternative scenario for 
the puzzling coexistence of charge fluctuations 
as well as the systematic change of the CO phase diagram in
$\theta$-(ET)$_2X$ by focusing on different roles of the off-site Coulomb repulsion and
the kinetic energy of electrons.

Our Hamiltonian consists of two contributions as
$
{\cal H} = {\cal H}_{\text{EHM}} + 
{\cal H}_{\text{el-ph}}
$.
The former term represents the extended Hubbard model in the form 
\begin{align}
{\cal H}_{\text{EHM}} & = t_p \! \! \! 
\sum\limits_{\langle i,j \rangle_p, \sigma} 
\! \! (c_{i\sigma}^{\dagger} c_{j\sigma} + \text{H.c.}) 
 + t_c \! \! \! \sum\limits_{\langle i,j \rangle_c, \sigma} 
\! \! (c_{i\sigma}^{\dagger} c_{j\sigma} + \text{H.c.}) \nonumber \\
&
 + U \sum\limits_i n_{i\uparrow} n_{i\downarrow} +
 V \sum\limits_{\langle i,j \rangle} n_i n_j.
\end{align}
Here, the summations with $\langle i,j \rangle_{p(c)}$ are taken over
the nearest-neighbor sites along the $x, y$ axes ($c$ axis) (Fig.~\ref{lattice}).
The off-site Coulomb interaction $V$ is taken to be isotropic 
because of the almost equal distances between neighboring ET molecules. 
We consider the case of $t_p, t_c > 0$ and
the average electron density at $1/4$ filling, 
which corresponds to $\theta$-(ET)$_2X$ by the electron-hole transformation. 
We take $t_p=1$ as an energy unit, and set the Boltzmann constant $k_{\text{B}} = 1$.

The latter term ${\cal H}_{\text{el-ph}}$ describes the electron-phonon couplings 
which are necessary because CO appears with a change of the lattice structure
\cite{rf:MoriH98_1,rf:MoriH98_2,rf:Clay02}.
Although realistic phonon modes are complicated, 
here we simply incorporate
the Holstein type and the Su-Shrieffer-Heeger (SSH) type phonons 
in the model as 
${\cal H}_{\text{el-ph}} = {\cal H}_{\text{Holstein}} + {\cal H}_{\text{SSH}}
$
with
\begin{align}
& {\cal H}_{\text{Holstein}} = 
g \sum\limits_{i} u_i n_i 
+ \sum\limits_i \Big\{ \frac{M \omega_0^2}{2} u_i^2 
+ \frac{1}{2M} \Pi_i^2 \Big\}, \\
& {\cal H}_{\text{SSH}} = g_p \! \! \! \sum\limits_{\langle i,j \rangle_p, \sigma} 
(u_{a,j} - u_{a,i})
(c_{i\sigma}^{\dagger} c_{j\sigma} + \text{H.c.}) \nonumber \\
& + g_c \! \! \! \sum\limits_{\langle i,j \rangle_c, \sigma}
(u_{c,j} - u_{c,i})
(c_{i\sigma}^{\dagger} c_{j\sigma} + \text{H.c.}) \nonumber \\
& + \sum\limits_i \Big\{ \frac{M' \omega_0'^2}{2} (u_{a,i}^2 + u_{c,i}^2)  +
 \frac{1}{2M'} (\Pi_{a,i}^2 + \Pi_{c,i}^2) \Big\}.
\end{align}
In ${\cal H}_{\text{Holstein}}$, $g$ is the coupling constant; 
$u_i$ and $\Pi_i$ denote the lattice distortion and
its conjugate momentum at site $i$, respectively. 
In ${\cal H}_{\text{SSH}}$, we describe the displacements of $i$th site 
in the $a$ and $c$ directions by $u_{a,i}$ and $u_{c,i}$, respectively, 
and we assume that they modulate the transfer integrals 
$t_p$ and $t_c$ independently
with the coupling constants $g_p$ and $g_c$; 
the summations are taken in a manner that
an elongation (shrinkage) of the bond length 
decreases (increases) the transfer integrals.
$\Pi_{a,i}$ and $\Pi_{c,i}$ are the conjugate momenta of
$u_{a,i}$ and $u_{c,i}$, respectively.
$M$, $M'$ and $\omega_0$, $\omega_0'$ are mass and eigen-frequency 
of the corresponding phonons, respectively.

To study CO phenomena in our model, 
we mainly analyze the charge susceptibility $\chi_{\text{c}}({\mathbf q}, i\omega_q)$ 
by RPA. 
For instance, in the case of the model with the Holstein phonons 
(${\cal H}_{\text{SSH}} = 0$),
$\chi_{\text{c}}$ is given in the form
\begin{equation}
\chi_{\text{c}}({\mathbf q}, i\omega_q) = \frac{\chi^{(0)}({\mathbf q}, i\omega_q)}
{1 + W({\mathbf q}, i\omega_q) \ \chi^{(0)}({\mathbf q}, i\omega_q)},
\label{rpa}
\end{equation}
where 
\begin{equation}
W({\mathbf q}, i\omega_q) = U + 2V({\mathbf q}) 
+ 2 \gamma^2 D^{(0)}({\mathbf q}, i\omega_q),
\label{Wq}
\end{equation}
and $\chi^{(0)}$ is the bare susceptibility given by 
$\chi^{(0)}({\mathbf q}, i\omega_q) = \frac{1}{N} \sum_{\mathbf p} 
\{ f(\varepsilon_{{\mathbf p}+{\mathbf q}}) - f({\varepsilon_{\mathbf p}}) \} /
\{ i\omega_q - (\varepsilon_{{\mathbf p}+{\mathbf q}} - \varepsilon_{\mathbf p}) \}$;
$N$ is the number of sites, 
$f$ is the Fermi distribution function, and 
$\varepsilon_{\mathbf p} = 2t_p (\cos p_x + \cos p_y) + 2t_c \cos (p_x + p_y)$.
Here, $V({\mathbf q})$ is the Fourier transform of the off-site Coulomb interaction,
$V({\mathbf q}) = 2V \{ \cos q_x + \cos q_y + \cos(q_x + q_y) \}$;
$D^{(0)}$ is the bare phonon Green's function 
given by 
$ D^{(0)}({\mathbf q}, i\omega_q) = - 2\omega_0 / ( \omega_q^2 + \omega_0^2) $,
and 
$ \gamma = g / \sqrt{2M\omega_0} $.
The expression becomes more complicated when ${\cal H}_{\text{SSH}}$ is included, 
but the main features of our results below are retained. 

\begin{figure}[t]
\begin{center}
\includegraphics[width=0.39\textwidth]{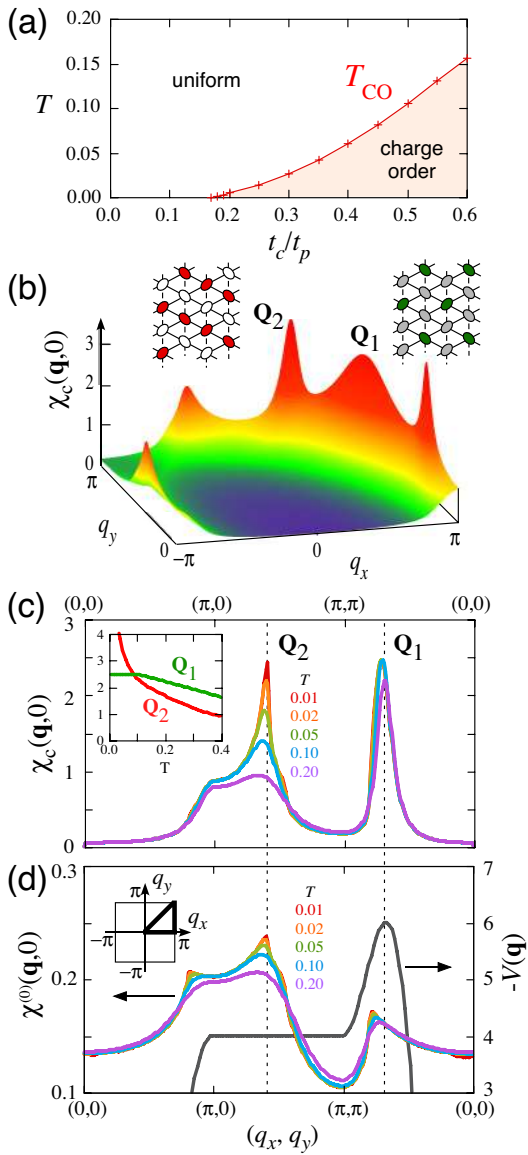}
\end{center}
\caption{\label{phase} (Color online).
(a) RPA phase diagram for the model with the Holstein phonons 
at $U=3.2$, $V=1.0$, and $\gamma^2/\omega_0=1.5$. 
(b) Charge susceptibility $\chi_{\text{c}}({\mathbf q}, 0)$ 
at $t_c/t_p=0.1$ and $T = 0.05$.
The insets show
CO patterns corresponding to the peak wave numbers.
(c) Temperature dependence of the charge susceptibility.
The inset shows
the peak heights.
(d) Fourier transform of the off-site Coulomb interaction $-V({\mathbf q})$,
and bare charge susceptibility $\chi^{(0)}({\mathbf q}, 0)$.
The inset shows the parametrization of ${\mathbf q} = (q_x, q_y)$ 
in the plots of (c) and (d).
}
\end{figure}

Now we discuss RPA results, mainly for the model 
without the SSH-type coupling (${\cal H}_{\text{SSH}} = 0$) for simplicity.
Figure~\ref{phase}(a) shows the phase diagram in the $t_c/t_p$-$T$ plane
for a typical set of parameters. 
$T_{\text{CO}}$ is determined by the divergence of the static charge susceptibility,
$\chi_{\text c} ({\mathbf q}, 0)$. 
The divergence appears at ${\mathbf q} = {\mathbf Q}_2 \sim (\pi, 0)$ [and $(0, \pi)$] 
although ${\mathbf Q}_2$ weakly depends on $t_c/t_p$.
${\mathbf Q}_2$ corresponds to a stripe-type CO pattern 
with period two along the $c$ axis 
as in the left inset of Fig.~\ref{phase}(b) 
(diagonal-type CO \cite{rf:Seo00}).
$T_{\text{CO}}$ is lowered as decreasing $t_c/t_p$ and
goes to zero at $t_c/t_p \sim 0.1$, a parameter corresponding to the Cs
compound in which $T_{\text{CO}} \to 0$ is experimentally observed. 
Here, we note that the Holstein phonon stabilizes CO 
by effectively reducing the on-site repulsion $U$
\cite{rf:note2}.

In the quantum critical regime at $t_c/t_p \sim 0.1$ where $T_{\text{CO}} \to 0$,
$\chi_{\text{c}} ({\mathbf q}, 0)$ shows two broad peaks at different wave numbers
as shown in Fig.~\ref{phase}(b):
One is at ${\mathbf Q}_1 \simeq (2\pi/3, 2\pi/3)$ and 
the other is at ${\mathbf Q}_2 \sim (\pi, 0)$ [and $(0, \pi)$].
Here the ${\mathbf Q}_1$ peak corresponds to CO pattern with period three,
as shown in the right inset of Fig.~\ref{phase}(b) 
\cite{rf:MoriT03}.
These two peaks exhibit contrastive $T$ dependences
as shown in Fig.~\ref{phase}(c):
The ${\mathbf Q}_1$ peak shows little $T$ dependence in this $T$ range, 
while the ${\mathbf Q}_2$ peak grows rapidly as $T$ decreases.
The behaviors are 
consistent with the X-ray scattering data in the Cs compound 
\cite{rf:WatanabeM99,rf:Nogami99}.

What is the origin of the coexistence of two charge fluctuations 
and their contrastive $T$ dependence?
The charge fluctuations at ${\mathbf Q}_1$ and ${\mathbf Q}_2$ 
have different origins, the off-site Coulomb interaction $V$ and 
the nesting property of the Fermi surface, respectively, as explained below.
From eqs.~(\ref{rpa}) and (\ref{Wq}), it is clear that 
$\chi_{\text{c}} ({\mathbf q}, 0)$ is enhanced at the wave numbers 
where the Fourier transform of the off-site Coulomb interaction
$V({\mathbf q})$ is minimized or the bare susceptibility $\chi^{(0)} ({\mathbf q}, 0)$ is maximized.
Figure~\ref{phase}(d) plots $-V({\mathbf q})$ and $\chi^{(0)}({\mathbf q}, 0)$.
Obviously, the ${\mathbf Q}_1$ peak corresponds to the minimum of $V({\mathbf q})$,
indicating that this peak originates from the off-site Coulomb repulsion. 
On the other hand, the ${\mathbf Q}_2$ peak corresponds to 
the maximum of $\chi^{(0)} ({\mathbf q}, 0)$, 
and hence this originates from 
the noninteracting electronic state, i.e., 
the nesting property of the Fermi surface. 
These assignments explain 
the contrastive $T$ dependence of the two peaks:
The ${\mathbf Q}_1$ peak is of the Coulomb origin, 
and hence its height is weakly dependent on $T$ 
in the present $T$ range much lower than the energy scale of $V$.
On the other hand, the Fermi degeneracy at low $T \ll t_p$ enhances 
$\chi^{(0)} ({\mathbf q}, 0)$ at the nesting vector 
${\mathbf q} = {\mathbf Q}_2$, 
which leads to the rapid increase of the ${\mathbf Q}_2$ peak. 
In the inset of Fig.~\ref{phase}(c), 
we show $T$ dependencies of the peak heights in a wider $T$ range.
The ${\mathbf Q}_2$ peak steeply grows at low $T$
with a concave $T$ dependence, 
while the ${\mathbf Q}_1$ peak has a weak, convex $T$ dependence and 
saturates at $T \sim 0.1$ 
\cite{rf:note}.

Thus, our RPA results successfully reproduce
the experimental features in $\theta$-(ET)$_2X$, i.e.,
the coexisting charge fluctuations and their contrastive $T$ dependences 
as well as the systematic change of phase diagram in terms of $t_c/t_p$. 
The coexistence of charge fluctuations 
comes from the fact that
the off-site Coulomb energy and the kinetic energy (the Fermi surface nesting) 
favor the different types of CO:
the former induces an instability toward the ``Wigner crystal'' formation, 
whereas the latter leads to the charge-density-wave instability.
The mechanism is intrinsic in the correlated electron systems, and it
holds for the model including ${\cal H}_{\text{SSH}}$. 

There remain some issues to be clarified.
One is the order of CO phase transition.
The transition is always of second order in RPA,
while it is of first order with a structural change in experiments
\cite{rf:MoriH98_1,rf:MoriH98_2}.
An interesting experimental observation is that 
a diffuse peak for the three-fold period appears above $T_{\text{CO}}$
\cite{rf:WatanabeM03,rf:WatanabeM04}
with an unusual critical behavior in spite of the first-order transition 
\cite{rf:Miyagawa00,rf:Chiba01}.
A corresponding behavior is found in our RPA results;
the threefold $\mathbf{Q}_1$ peak develops above $T_{\text{CO}}$ 
since the peak is induced by $V$ much larger energy scale than $T_{\text{CO}}$. 
However, in order to describe the first-order nature of the transition, 
it is necessary to go beyond RPA.
Anharmonic contributions from phonons may play a role as well.
This is left for future study.

Another issue is the wave numbers of CO and fluctuations.
In our results, 
the coexisting charge fluctuations appear at
${\mathbf Q}_1 \simeq (2\pi/3, 2\pi/3)$ and 
${\mathbf Q}_2 \sim (\pi, 0)$. 
Whereas, in experiments, they are found at
${\mathbf q}_1 \simeq (\pi, -\pi/3)$ and
${\mathbf q}_2 \simeq (\pi/2, \pi/2)$ 
\cite{rf:WatanabeM99,rf:Nogami99}. 
Accordingly, the stabilized CO pattern below $T_{\text{CO}}$ is also different;
our result in Fig.~\ref{phase}(a) predicts 
the diagonal-type CO with ${\mathbf Q}_2 \sim (\pi,0)$, 
while the experimental one is the horizontal type 
with ${\mathbf q}_2 \simeq (\pi/2, \pi/2)$ 
\cite{rf:Seo00,rf:WatanabeM03,rf:WatanabeM04}.
We consider that these discrepancies are not fundamental
and can be reconciled without modifying the essential part of our results
in the following reasons.
As to the ${\mathbf Q}_1$ peak, the wave number is determined by 
the minimum of $V({\mathbf q})$, 
and it locates at ${\mathbf q}={\mathbf Q}_1$
because we assume the simplest form of the off-site Coulomb interaction,
i.e., the isotropic one for the nearest-neighbor sites only.
If we include longer-range parts of and/or an anisotropy in the Coulomb interaction, 
$V({\mathbf q})$ takes its minimum at a different wave number, 
which may coincide with ${\mathbf q}_1$ 
\cite{rf:Kuroki06}. 
On the other hand, the ${\mathbf Q}_2$ peak comes from 
the maximum of $\chi^{(0)}$, 
and hence, its wave number is dependent on the form of transfer integrals
which modifies the Fermi surface nesting. 
In fact, we observe a shift of the peak when we include the SSH-type coupling
which directly couples to the transfer integrals.

\begin{figure}[t]
\begin{center}
\includegraphics[width=0.45\textwidth]{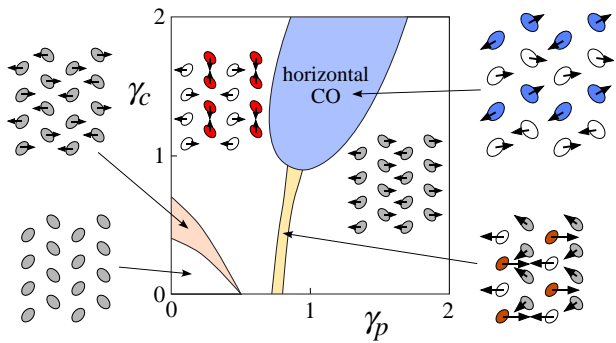}
\end{center}
\caption{\label{ssh_phase} (Color online).
Phase diagram for the model 
with the SSH phonons at $T=0.01$ and $U=V=\gamma=0$.
CO patterns and displacements of ET molecules 
are also shown.}
\end{figure}

To demonstrate how the SSH-type coupling affects the ordering vector ${\mathbf Q}_2$, 
we study the low-$T$ phase diagram 
for the model including ${\cal H}_{\text{SSH}}$.
We decouple Coulomb interactions and electron-phonon couplings 
by using the mean fields,
$\langle c_{i\sigma}^\dagger c_{i\sigma} \rangle$,
$\langle c_{i\sigma}^\dagger c_{j\sigma} \rangle$,
$\langle u_i \rangle$ and $\langle u_{a(c),i} \rangle$, and
obtain a solution by minimizing the free energy 
among various mean-field configurations 
within a unit cell of up to $4 \times 4$ sites in the $x$ and $y$ coordinates. 
Figure~\ref{ssh_phase} shows the result at $T=0.01$
by changing the SSH couplings $\gamma_p = g_p / \sqrt{2M'\omega_0'}$ and
$\gamma_c = g_c / \sqrt{2M'\omega_0'}$.
To focus on the effect of ${\cal H}_{\text{SSH}}$, 
here we omit the Holstein coupling and take $U=V=0$.
A variety of phases emerges as shown in the figure; 
in particular, 
in a region where both $\gamma_p$ and $\gamma_c$ are substantial, 
the horizontal-type CO is stabilized 
whose wave vector is
consistent with ${\mathbf q}_2$ observed in Tl and Rb compounds
\cite{note3}.
We confirmed that the horizontal CO remains robust against finite $U$
and $V$ within the Hartree-Fock approximation.
Thus, the SSH-type electron-phonon coupling plays an important role in
determining an actual CO instability induced, 
which suggests possibility to shift ${\mathbf Q}_2$ to ${\mathbf q}_2$.
For further comparison,
detailed information is indispensable on realistic electron-phonon couplings, 
which are 
possibly more complicated than our model.

In summary, 
we have theoretically investigated the origin of puzzling features 
in charge ordering and fluctuations in $\theta$-(BEDT-TTF)$_2X$.
The peculiar coexisting fluctuations are ascribed to 
two different instabilities,
the ``Wigner crystallization'' driven by 
the Coulomb repulsions and 
the charge-density-wave instability due to the Fermi surface nesting.
Our scenario does not need to suppose 
either a competition among different ``Wigner-crystal-type'' CO
in the strong-correlation regime 
nor real-space inhomogeneity. 
Our results provide a key toward understanding 
the low-$T$ transport phenomena related to the charge fluctuations. 
 
The authors thank M.\ Imada, H.\ Seo, S.\ Fujiyama, K.\ Miyagawa, 
Y.\ Nogami, I.\ Terasaki, Y.\ Mutoh, T.\ Kato, N.\ Furukawa, and C.\ Hotta for fruitful discussions. 
This work was supported by Grants-in-Aid for Scientific Research 
(Nos. 17071003 and 16GS50219) from the MEXT.
One of the authors (M. U.) is supported by JSPS.


\begin{thebibliography}{9}
\bibitem{rf:note} 
Note that the ${\mathbf Q}_2$ peak value is larger than that in
the main panel of Fig.~\ref{phase}(c) 
because the peak is slightly off from the Brillouin zone boundary.
\bibitem{note3}
The origin of the horizontal CO appears to be different 
from that in Ref.~\cite{rf:Clay02} 
since our state is stabilized by the SSH phonons and
appears even at $U=V=0$.
\end{thebibliography}
\end{document}